\documentstyle[12pt]{article}
\begin{document}

\begin{center}
\hfill MKPH-T-94-18\\
\vspace*{3.3cm}
{THE DRELL-HEARN-GERASIMOV SUM RULE\footnote{Supported by the
Deutsche Forschungsgemeinschaft (SFB 201)}}\\[0.8cm]
D. DRECHSEL\\[0.4cm]
Institut f\"ur Kernphysik, Universit\"at Mainz\\
D-55099 Mainz, Germany\\
\end{center}

\vspace{2.0cm}
\hspace{2.5cm}{\small ABSTRACT\\[2ex]
The Drell-Hearn-Gerasimov (DHG) sum rule relates the helicity structure of the
photoabsorption cross section to the anomalous magnetic moment of the nucleon.
It is based on Lorentz and gauge invariance, crossing symmetry, causality and
unitarity. A generalized DHG sum rule my be derived for virtual photons. At low
momentum transfer this generalized sum rule is saturated by the resonance
region,  at high momentum transfer it may be expressed by the parton spin
distributions measured in deep inelastic scattering. The
longitudinal-transverse
interference determines the Cottingham sum rule, which is related to the
electric and magnetic form factors over the whole range of momentum tranfer.
\\[2ex]

\hspace{2.5cm}KEYWORDS\\[2ex]
Photoabsorption, sum rules, helicity, asymmetries, spin structure, quark model}
\\[2ex]

\hspace{2.5cm}INTRODUCTION\\[2ex]
The existence of internal degrees of freedom manifests itself in a finite size
of the nucleon, described by a form factor of a Dirac current and an anomalous
magnetic moment multiplied by the Pauli form factor. By the same token a
spectrum of excited states appears, a series of resonances in the mass region
of
1-2 $GeV$ and a flat continuum at higher energies, logarithmically rising at
the
highest observed energies between 200 - 300 $GeV$. Finite size effects in the
ground state and the existence of an excitation spectrum are not all
independent
phenomena, but closely intertwined by sum rules and low energy theorems (LET).

On the experimental side, photo- and electronuclear reactions are a
particularly
clean instrument to investigate the resonance region and to analyze the
multipole content of the individual resonance contributions.
With the advent of electron accelerators of high current and large duty-factor,
new classes of experiments including polarization degrees of freedom have
become
possible. Such investigations range from threshold production of mesons to
detailed studies of the helicity structure in the resonance region. The
helicity
structure of the cross section is expected to change at momentum transfers of
the order of the vector meson masses.

The Drell-Hearn-Gerasimov (DHG) and Burkhardt-Cottingham (BC) sum rules connect
the helicity structure of the cross sections in the inelastic region with
ground state properties. Being based on general principles of physics like
Lorentz and gauge invariance, crossing symmetry, causality and unitarity, these
sum rules are an important consistency check for our understanding of the
hadronic structure. They have never been measured directly. However, an
analysis
of pion photoproduction indicates some problems with the proton-neutron
difference for the DHG sum rule. New experiments are underway to investigate
these questions. Of particular interest is the question whether and how fast
these sum rules converge as functions of the excitation energy. A failure to
converge would shed serious doubts on our present understanding of hadronic
structure and send the model-builders back to the drawing board.

As function of momentum transfer $Q^2$, chiral perturbation theory (ChPT)
predicts the slope of the DHG integral at the real photon point. However, the
loop expansion of ChPT breaks down in the region of the vector meson
resonances,
where the helicity structure changes abruptly. Similarly, we have solid
predictions for $Q^2\rightarrow\infty$ from perturbative $QCD$. In the scaling
region the DHG and BC sum rules may be directly expressed by the spin
distribution functions of the quarks, the object of deep inelastic lepton
scattering. Again, perturbative $QCD$ breaks down if we approach the region of
the vector meson masses, now from above. Corresponding to the pole structure in
the complex plane, the resonance region will define a circle of convergence for
both
an expansion at the origin (the loops of ChPT) and at infinity (higher twists
of
perturbative QCD).

In the following sect. 2 we will discuss the "classical" DHG sum rule for real
photons. The more general framework of electroproduction including polarization
degrees of freedom will be outlined in sect. 3. Appropriately defined
integrated
cross sections yield a generalization of the DHG sum rule to virtual photons
and, derived from the longitudinal-transverse interference, the BC sum rule.
Theories and models for these sum rules will be presented in sect. 4. Finally,
we will briefly review the existing information on the helicity structure of
the
low-lying resonances in sect. 5, and draw some conclusions in sect. 6.
\\[2ex]

\hspace{2.5cm}THE DHG SUM RULE FOR REAL PHOTONS\\[2ex]
The differential cross section for Compton scattering off the nucleon
(for the kinematics see Fig. 1)
may be decomposed into the contribution of the point-like
Dirac particle as evaluated by Klein and Nishina (1929),
additional contributions of the anomalous magnetic moment $\kappa$ as given by
Powell (1949) and terms arising from virtual excitations, e.g. the
polarizabilities of the nucleon,
\begin{equation}
\frac{d \sigma}{d \Omega} = \left. \frac{d \sigma}{d \Omega} \right|_{KN}
+ \left. \frac{d \sigma}{d \Omega} \right|_P ( \kappa, \kappa^2, \kappa^3,
\kappa^4 )
+ \left. \frac{d \sigma}{d \Omega} \right|_{pol}.
\end{equation}
In the case of forward scattering (photon scattering angle $\theta =0$),
only the terms
of quartic order in $\kappa$ remain finite. The corresponding scattering
amplitude,
\begin{equation}
T ( \omega, \theta= 0 ) = {\hat \epsilon}'^{*} \cdot {\hat \epsilon }
f ( \omega ) + i \vec \sigma \cdot ( {\hat \epsilon}'^{*}
\times {\hat \epsilon} ) g ( \omega ),
\end{equation}
contains a spin-flip amplitude $g$ and a no-flip amplitude $f$,
both functions of
the photon energy $\omega$. The polarization vectors of the initial and final
photon are denoted by $\epsilon$ and $\epsilon'$, respectively, and
$\vec \sigma$ is the spin of the nucleon.

The amplitudes $f$ and $g$ may be expanded into a power series in $\omega$
whose
leading terms are determined by low energy theorems (LET) based on relativity
and
gauge invariance (Low, 1954; Gell-Mann and Goldberger, 1954)
\begin{eqnarray}
f ( \omega ) & = & - \frac{e^2}{m} + ( \alpha + \beta ) \omega^2 + [\omega^4],
\\
g ( \omega ) & = & - \frac{e^2 \kappa^2}{2 m^2} \omega + \gamma \omega^3 +
[\omega^5].
\end{eqnarray}
The leading term in $f$ is the famous Thomson limit, the next order term is the
contribution of the scalar polarizabilities of the nucleon, a sum of electric
($\alpha$) and magnetic ($\beta$) terms. The leading term in the spin-flip
amplitude is
proportional to the square of the anomalous magnetic moment; the next order
term
is the vector polarizability. The low energy limit of $g$ is due to a Feynman
graph with $\kappa$
operating at both $\gamma NN$ vertices, leading to the ${\kappa}^4$
contribution in
the total cross section.

The two terms $f$ and $g$ may be separated by an experiment using circularly
polarized photons and nucleons polarized with spin parallel or antiparallel to
the photon momentum. As has been shown in Fig. 2, the former situation leads to
a state with overall spin $J_z = \frac{3}{2}$, the latter process to $J_z =
\frac{1}{2}$. The corresponding amplitudes may
be evaluated using eq. (2),
\begin{equation}
T_{3/2} = f-g,\qquad T_{1/2} = f+g.
\end{equation}
The optical theorem relates the imaginary parts of these amplitudes to the
corresponding total absorption cross sections,
\begin{equation}
\mbox{Im}\, T_{1/2,3/2}(\omega) = \frac{\omega}{4\pi} \sigma_{1/2,3/2}(\omega).
\end{equation}
Furthermore $f$ is an even and $g$ an odd function under
$\omega\rightarrow\ -\omega$ (crossing symmetry). On the basis of analyticity,
unitarity
and crossing symmetry, we may write a dispersion relation for g,
\begin{equation}
\mbox{Re}\, g(\omega) = \frac{2\omega}{\pi} \int_{thr}
^\infty\frac{d\omega'}{\omega'^2
-\omega^2} \frac{\omega'}{4\pi} \frac{\sigma_{1/2}-\sigma_{3/2}}{2}\,.
\end{equation}

Since the threshold energy is of the order of the mass of the pion, $m_{\pi}$,
this
expression may be expanded into a power series in $\omega$. Comparing this
series with the low energy expansion, eq. (4), we obtain the DHG sum rule
(Drell and  Hearn, 1966; Gerasimov, 1966)
\begin{equation}
-\frac{\kappa^2}{4} = \frac{m^2}{8\pi^2\alpha}\int^\infty\frac{d\nu}{\nu}
(\sigma_{1/2}(\nu)- \sigma_{3/2}(\nu))\stackrel{!}{=} I(Q^2 = 0)\,.
\end{equation}
Here and in the following we denote
\begin{eqnarray}
\nu &=& \frac{p\cdot q}{m} = \omega_{lab}\\
Q^2 &=& - q^2 =\left\{\begin{array}{r@{\quad:\quad}l} 0 &\mbox{real photons}\\
>0 &\mbox{electron scattering}\,,\end{array}\right.\nonumber
\end{eqnarray}
and $\alpha = e^2/4\pi\approx$ 1/137. On the $rhs$ of eq. (8) we have defined
the
real photon point of a function $I (Q^2)$ whose meaning will become clear in
the following section. Similar to eq. (8) also the vector polarizability (or
higher moments!) may be related to sum rules,
\begin{equation}
\gamma = \frac{1}{4\pi^2} \int\frac{d\nu}{\nu^3} (\sigma_{1/2} - \sigma_{3/2}).
\end{equation}
In the more general formalism of photoabsorption (or electron scattering, see
sect. 3), the helicity cross sections are related to the total transverse
$(\sigma_T)$ and "transverse-transverse" $(\sigma_{TT'})$ cross sections,
\begin{eqnarray}
\sigma_T &=& \frac{\sigma_{3/2}+\sigma_{1/2}}{2}\\
\sigma_{TT'}&=& \frac{\sigma_{3/2} - \sigma_{1/2}}{2}\,.
\end{eqnarray}
Our experimental knowledge about these quantities is summarized in Figs. 3 and
4. The cross section $\sigma_T$ clearly shows the first and second resonance
region, indications of two more
broad peaks and a nearly constant value to
energies of about $180 GeV$. The more recent DESY data lead up to the order of
$300 GeV$ and show a slow logarithmic increase. As a consequence a dispersion
relation for the Thomson term would not converge,  and only a once-subtracted
dispersion relation can be established for the sum of the scalar
polarizabilities, $\alpha+\beta$. Being the difference of the helicity cross
sections, $\sigma_{TT'}$ is expected to decrease slowly with $\nu$, which would
guarantee the convergence of the DHG integral, eq. (8).

The DHG sum rule has never been measured directly, the results shown in Fig.4
are essentially based
on phase shift analyses of pion photoproduction using some estimates for the
two-pion background. It
involves data on both the proton and the neutron, because $\kappa^2$ has
the isospin dependence
\begin{equation}
\kappa^2 = (\kappa_S + \tau_0 \kappa_V )^2
= \kappa_V^2 + \kappa_S^2 + 2 \kappa_S \kappa_V \tau_0.
\end{equation}
Obviously the DHG integral $I$ is dominated by the isovector moment
$\kappa_V$ (term $I_{VV}$),
the proton-neutron difference ($I_{SV}$)
is smaller by an order of magnitude and
the contribution of the isoscalar moment ($I_{SS}$) is practically negligible.
The
experimental data show clear indications for resonance structures with
oscillating sign of the integrand of $I$ (see Fig. 4). A more detailed
multipole
decomposition of $I$ is given in table 1. It shows good agreement
between experiment and the sum rule prediction for $I_{VV}$, but a large
discrepancy for $I_{SV}$. This has led to the speculation that the latter
integral might need a
subtraction. Chang {\it et al.} (1992) have tried to reconcile experiment and
theory within the framework of a generalized current algebra. However, the
paper
has never been published.\\[1ex]
\begin{table}[htbp]
{\small Table 1:
The multipole structure of the DHG integral $I_p (Q^2 = 0)$ for the 3 isospin
channels VV, SV and SS (see text).
For the definitions of the resonances and multipoles see (Drechsel and Tiator,
1992). The "experimental" numbers are obtained by an analysis of pion
photoproduction
(Karliner, 1973).}
\begin{center}
\begin{tabular*}{12cm}{l@{\extracolsep{\fill}}lrrc}
resonance & multipole & $I^{\mbox{vv}}$ & $I^{\mbox{sv}}$ & $I^{\mbox{ss}}$ \\
\hline
$P_{33}, 3/2^+$ & $M_{1+} (E_{1+})$ & -1.05 & +.09 & -\\
\multicolumn{2}{r}{$(\Delta_{1232}$ only}& -.93 &\multicolumn{1}{l}{)} &\\
$S_{11}, 1/2^-$ & $E_{0+} $         & +.65  & -.09 & -\\
$P_{11}, 1/2^+$ & $M_{1-} $         & +.04  & -.01 & -\\
$D_{13}, 3/2^-$ & $E_{2-}, M_{2-}$  & -.26  & -.05 & -\\
$F_{15}, 5/2^+$ & $E_{3-}, M_{3-}$  & -.04  & -.03 & -\\
\multicolumn{2}{r}{2$\pi$ background}& -.20 & -.06 & -\\ \hline
\multicolumn{2}{r}{experiment}      & -.86  & -.15 & small\\
\multicolumn{2}{r}{DHG}             & -.86  & +.06 & -.001\\ \hline
\end{tabular*}
\end{center}
\end{table}

\vspace{2ex}

\hspace{2.5cm} THE GENERALIZED DHG FOR ELECTRON SCATTERING\\[2ex]
The kinematics of lepton scattering with polarization degrees of freedom is
shown in Fig. 5a for target polarization. The (longitudinal) polarization of
the
high energy electron is denoted by
$h=\vec{\sigma}\cdot\hat{k}\rightarrow\pm 1$, the polarization $\vec{P}$ of the
target nucleon may be decomposed into a
coordinate system with ${\hat e}_z = {\hat q} $,
along the direction of the virtual photon,
$ {\hat e}_x \perp {\hat e}_z $ in the electron scattering
plane and in the hemisphere of the outgoing electron, and
$ {\hat e}_y = {\hat e}_z \times {\hat e}_x $ perpendicular to
the scattering plane. Note that in the standard EMC/SLAC experiment,
${\hat P} = \pm {\hat k}$,
e.g. the spins of nucleon and electron are parallel or antiparallel.

In the case of a coincidence experiment, e.g. $ e + p \rightarrow e' + p' + \pi
$,
the recoil polarization is usually analyzed in a coordinate system connected
with the reaction plane of the
$p'-\pi$ system. Its axes are
denoted by ${\hat l}$ (along the direction of $p'$), ${\hat t}$
(transverse, in the reaction plane) and ${\hat n}$
(normal to the reaction plane), as
shown in Fig. 5b. The cross section for such an experiment is given by
(Drechsel and Tiator, 1992)
\begin{equation}
\frac{d \sigma}{d \Omega_{e'} d k_{e'} d \Omega_{\pi}} =
\Gamma \frac{d \sigma^{(v)}}{d \Omega_{\pi}},
\end{equation}
where $\Gamma$ is the flux and $d \sigma^{(v)} / d \Omega_{\pi}$
the differential cross section for the
virtual photon,

\newpage
\begin{eqnarray}
\frac{d \sigma^{(v)}}{d \Omega_{\pi}} & =
{\displaystyle \frac{\mid \vec k_{\pi} \mid}{k_{\gamma}^{cm}} }
\Big\{& R_T + P_n R_T^n
\nonumber \\
& &+ \varepsilon_L ( R_L + P_n R_L^n )
\nonumber \\
& &+ \sqrt{2 \varepsilon_L (1 + \varepsilon)}
\left( ( R_{TL} + P_n R_{TL}^n) \cos \Phi
+ (P_l R_{TL}^l + P_t R_{TL}^t) \sin \Phi \right)
\nonumber \\
& &+ \varepsilon
\left( ( R_{TT} + P_n R_{TT}^n) \cos 2 \Phi
+ (P_l R_{TT}^l + P_t R_{TT}^t) \sin 2 \Phi \right)
\nonumber \\
& &+ h \sqrt{2 \varepsilon_L ( 1 - \varepsilon )}
\left( ( R_{TL'} + P_n R_{TL'}^n) \sin \Phi
+ ( P_l R_{TL'}^l + P_t R_{TL'}^t ) \cos \Phi \right)
\nonumber \\
& &+ h \sqrt{1 - \varepsilon^2} (P_l R_{TT'}^l + P_t R_{TT'}^t)
\,\Bigr\},
\end{eqnarray}
with $\varepsilon$ and $\varepsilon_L$
the transverse and "longitudinal"polarizations of the virtual
photon, $k_{\gamma}^{cm}$ the "photon equivalent energy" in the $cm$ frame
(Drechsel and Tiator, 1992), and
all quantities being expressed in that frame. For an inclusive reaction the
cross section has to be summed over the azimuthal angle $\Phi\equiv
\Phi_{\pi}$.
Due to their
definition with regard to the reaction plane, also the components ${P_l, P_t,
P_n}$ depend on the pion angles, $(\Theta_{\pi}, \Phi_{\pi})$.
In this way also combinations like $P_n \sin \Phi$, etc.,
give finite contributions to the angular integration. Defining then the
inclusive cross sections (in a somewhat symbolic way !) by
\begin{equation}
\sigma_i = \int \frac{\mid \vec k_{\pi} \mid}{k_{\gamma}^{cm}} R_i (
\Theta_{\pi} ) d \Omega_{\pi},
\end{equation}
we obtain
\begin{equation}
\sigma^{(v)} = \sigma_T + \varepsilon_L \sigma_L + h P_x
\sqrt{2 \varepsilon_L ( 1 - \varepsilon )} \sigma_{LT'}
+ h P_z \sqrt{1 - \varepsilon^2} \sigma_{TT'},
\end{equation}
i.e. two structure functions $(L,T)$ without polarization and two others $(LT'$
and
$TT')$ for a double polarization experiment. Up to kinematical factors, the
four
partial cross sections may be expressed by the CGLN multipoles (Chew {\it et
al.}, 1957),
\begin{eqnarray}
\sigma_T & = & 4\pi \frac{\mid \vec k_{\pi} \mid}{k_{\gamma}^{cm}} \sum_l
\frac{1}{2} ( l + 1 )^2
\left[(l+2)(\mid E_{l+} \mid^2 + \mid M_{l+1,-} \mid^2)
+ l ( \mid M_{l+} \mid^2 + \mid E_{l+1,-} \mid^2) \right], \\
\sigma_L & = & 4\pi \frac{\mid \vec k_{\pi} \mid}{k_{\gamma}^{cm}}\sum_l ( l +
1 )^3
\left[ \mid L_{l+} \mid^2 + \mid L_{l+1,-} \mid^2 \right], \\
\sigma_{LT'} & = & 4\pi \frac{\mid \vec k_{\pi}
\mid}{k_{\gamma}^{cm}}\sum_l\frac{1}{2}( l + 1 )^2
\left[ - L_{l+}^* \left( ( l + 2 ) E_{l+} + l M_{l+} \right) +
L_{l+1,-}^* \left( l E_{l+1,-} + ( l + 2 ) M_{l+1,-} \right) \right], \\
\sigma_{TT'} & = & 4\pi \frac{\mid \vec k_{\pi} \mid}{k_{\gamma}^{cm}}\sum_l
\frac{1}{2} ( l + 1)
\left[ - ( l + 2 ) \left( \mid E_{l+} \mid^2 + \mid M_{l+1,-} \mid^2 \right)
+ l \left( \mid M_{l+} \mid^2 + \mid E_{l+1,-} \mid^2
\right) \right. \nonumber \\
& & \left.\hspace{2.0cm} - 2 l ( l + 2 ) \left(
E_{l+}^* M_{l+} - E_{l+1,-}^* M_{l+1,-} \right) \right].
\end{eqnarray}
Note that the "unpolarized" functions, $\sigma_L$ and $ \sigma_T$,
contain only positive contributions,
while the "polarized" ones, $\sigma_{LT'}$ and $\sigma_{TT'}$,
are arithmetic sums with alternating signs.
The Fermi-Watson theorem (Watson, 1954) guarantees that phases of all
multipoles
with index $l_+$ or $l_-$ carry the same phase of the corresponding $\pi N$
partial
wave, i.e. $\sigma_{LT'}$ contains only a real part. It is interesting to note
that the
third line in eq. (15) will formally give rise to a cross section
$\sigma_{LT}$.
However,
this cross section is precisely  the imaginary part of the multipole
combination
of $\sigma_{LT'}$, i.e. it vanishes in the one-photon exchange approximation,
also in the
energy region of more-pion and other particle production because of the
unitarity  fo the S matrix.

The leading contributions for the two spin-polarized structure functions are
\begin{eqnarray}
\sigma_{LT'} & = & 4\pi \frac{\mid \vec{k}_{\pi} \mid}{k_{\gamma}^{cm}}
\{-2L^*_{1+} (M_{1+} + 3 E_{1+})+ L^*_{1-} M_{1-}-L^*_{0+} E_{0+}+L^*_{2-}
E_{2-}\,\pm\}\\
\sigma_{TT'} & = & 4\pi \frac{\mid \vec{k}_{\pi} \mid}{k_{\gamma}^{cm}}
\{\mid
M_{1+}\mid^2 - 6 E^*_{1+} M_{1+} - 3 \mid E_{1+}
\mid^2 - \mid M_1- \mid^2 - \mid E_{0+} \mid^2 + \mid E_{2-} \mid^2 \pm \}.
\end{eqnarray}
The bulk contribution to the DHG integrand, $\sigma_{TT'}$,
comes from the $\Delta (1232)$
resonance multipoles $1^+$; the higher resonances $N^*(1440), N^*(1535)$
including the S-wave threshold production, and $N^*(1520)$ contribute the
multipoles $1^-$,$0^+$ and $2^-$, in that order.
Finally, $\sigma_{TT'}$ and $\sigma_{LT'}$ are related to the Bjorken structure
functions (Bjorken, 1966) $G_1$ and $G_2$ by
\begin{eqnarray}
\sigma_{LT'} & = & - \frac{4 \pi^2 \alpha m}{1 - x} \left( G_1 ( \nu, Q^2) +
\frac{\nu}{m} G_2 (\nu, Q^2)
\right), \\
\sigma_{TT'} & = & - \frac{4 \pi^2 \alpha m}{1 - x} \left( G_1 ( \nu, Q^2) - 2
x
G_2 ( \nu, Q^2) \right),
\end{eqnarray}
where $x= Q^2/2m\nu$ is the Bjorken scaling variable. In terms of such
structure
functions, the inclusive cross section,
\begin{equation}
\frac{d^2 \sigma}{d \Omega_{e}' d k_{e}'} = \frac{\alpha^2}{Q^4}
\frac{k'}{k}
L_{\mu \nu} ( k ) W^{\mu \nu} ( p ),
\end{equation}
may be expressed by a leptonic $(L_{\mu\nu})$ and an hadronic tensor
$(W^{\mu\nu})$.
Both may be decomposed into a symmetric ${(\cal S)}$ and an antisymmetric
${(\cal A)}$
part,
\begin{eqnarray}
W_{\mu\nu}^{(\cal S)} &=& \frac{1}{m} \left(\frac{q_\mu q_\nu}{q^2}
- g_{\mu\nu} \right) F_1 ( \nu , Q^2 ) + \frac{1}{m \nu} \tilde p_{\mu}
\tilde p_{\nu} F_2 ( \nu , Q^2 ) \\
W_{\mu\nu}^{(\cal A)} &=& \epsilon_{\mu \nu \alpha \beta} q^\alpha
\left( m s^\beta G_1 + \frac{1}{m} \left( s^\beta p \cdot q - p^\beta
s \cdot q \right) G_2 \right), \nonumber
\end{eqnarray}
where $\tilde p^\nu = p^\nu - ( p \cdot q / q^2 ) q^\nu$ is a gauge
invariant vector. Except for normalization factors, the
leptonic tensor has the same structure as eq. (27),
with all form factors equal to
1. Since $ s^\beta \sim p ^\beta$ for high energetic leptons, only the first
term in $L_{\mu\nu}^{(\cal A)}$ contributes. Its contraction with
$W_{\mu\nu}^{(\cal A)} $ gives rise to the
spin-dependent parts of the cross section depending on the helicity $h$ of the
electron.

The DHG integral, eq. (8), is
\begin{eqnarray}
I (Q^2) = \frac{m^2}{4\pi\alpha} \int \frac{d\nu}{\nu} (1-x)
\sigma_{TT'} (\nu, Q^2),
\end{eqnarray}
and a similar integral, $J(Q^2)$, may be obtained for the structure function
$\sigma_{LT'}$. Note that the integral runs from threshold to infinity.
In the scaling region ( $Q^2, \nu \rightarrow \infty$; $x$ fixed ) the
structure
functions may be expressed by the quark distribution functions,
\begin{eqnarray}
m^2 \nu G_1 ( \nu, Q^2 ) &=& g_1 (x, Q^2) \Rightarrow
g_1(x) = \frac{1}{2} \sum_i e_i^2 \left( f_i^\uparrow (x)
-f_i^\downarrow (x)  \right) \\
m\nu^2 G_2 ( \nu, Q^2 ) &=& g_2 (x, Q^2) \Rightarrow
g_2(x) = \frac{1}{2} \sum_i e_i^2 \left( f_i^\rightarrow(x)
-f_i^\leftarrow(x)  \right) - g_1(x), \nonumber
\end{eqnarray}
where $f_i^{\uparrow,\downarrow}$ and $f_i^{\rightarrow,\leftarrow}$ denote the
densities for longitudinal and transverse quark polarization, respectively.
Changing the integration variable from $\nu$ to $x$, the two sum rules
may be expressed by the quark spin distributions,
\begin{eqnarray}
I ( Q^2 ) &=& \frac{ 2 m^2}{Q^2} \int dx \left[ g_1 ( x, Q^2)
- \frac{4 x^2 m^2}{Q^2} g_2 ( x, Q^2) \right],
\end{eqnarray}
\begin{eqnarray}
J ( Q^2 ) &=& \frac{ 2 m^2}{Q^2} \int dx \left[ g_1 ( x, Q^2)
+ g_2 ( x, Q^2) \right] \equiv J_1 + J_2 .
\end{eqnarray}
Since the second term in eq. (30) is small under the usual experimental
conditions, we expect
\begin{equation}
J_1(0) \approx I (0) = -\frac{ \kappa^2}{4} .
\end{equation}
As has been recently pointed out by Soffer and Teryaev (1993), the integral
$J_2$ is related to the so-called "Burkhardt-Cottingham sum rule" (Burkhardt
and Cottingham, 1970; Heimann, 1973).
Further aspects of this sum rule have been discussed in the early
70's (Feynman, 1972; Schwinger, 1975; Tsai {\it et al.}, 1975).
Since $g_2 (\nu, Q^2)$ is an odd function under
crossing, $(\nu \rightarrow - \nu)$, its sum over all
intermediate states vanishes.
As a consequence the contribution over the excited states is exactly cancelled
by the ground state expectation value, leading to
\begin{equation}
J_2 ( Q^2 ) = \frac{\mu}{4} G_M ( Q^2 ) \left( \mu G_M ( Q^2)
- G_E (Q^2 ) \right),
\end{equation}
where $\mu$ is the total magnetic moment, and $G_E$ and $G_M$
are the electric and magnetic form factors of the nucleon. In particular
\begin{equation}
J_2 ( 0) = \frac{1}{4} \kappa\mu , \qquad J (0) = \frac{1}{4}
\kappa(\mu-\kappa).
\end{equation}
\\[1ex]

\hspace{2.5cm} THEORIES AND MODELS

\underline{Experimental Status}\\[2ex]
The present "experimental" situation for the DGH sum rule for the proton
$(I_p)$
and neutron $(I_n)$ is summarized in Figs. 6 and 7, respectively, by the solid
line labeled "phenomenological model" (Burkert {\it et al.}, 1991; Kuhn {\it et
al.},
1993). It is obtained by
fitting a set of resonances, based on a relativistic quark model, to the data.
At $Q^2 = 0$ it agrees reasonably well with the previous analysis of pion
photoproduction (Karliner, 1973) and a later analysis by the Virginia group
(Workman and Arndt, 1992). The clear disagreement of these results with the DHG
prediction for
the neutron, as seen in Fig. 7, is certainly a good motivation to repeat the
experiment. Another striking feature is the rapid decrease from the large
absolute values at small $Q^2$ to values around zero at $Q^2\approx 1 GeV^2$.
Finally, for
$Q^2 \ge 2 GeV^2$, the DHG integral should have its asymptotic $Q^{-2}$
behaviour
with a constant determined by the EMC/SLAC experiments of deep inelastic
scattering (DIS).
The error bars given in the two figures indicate the projected
range and accuracy of the planned CEBAF experiments.\\[2ex]
\underline{Vector Meson Dominance (VMD)}\\[2ex]
A global fit to the data has been given in a model inspired by VMD
(Anselmino {\it et al.}, 1989),
\begin{equation}
I ( Q^2 ) = \left( -\frac{\kappa^2}{4} + \frac{{\cal{Z}} Q^2 m^2 }{m_V^2}
\right) \left/ \left( 1 + \frac{Q^2}{m_V^2} \right)^2 \right. ,
\end{equation}
where $m_V$ is the mass of the vector mesons and
$\cal Z $ has been determined by DIS.
It describes both the behaviour at small and large $Q^2$ and predicts at sign
change at $Q^2 \approx m_V^2$. In a somewhat different parametrization, Burkert
and Ioffe (1992) have fitted the sum rule to the $\Delta (1232)$
contribution plus monopole and dipole forms.\\[2ex]
\underline{Constituent Quark Model (CQM)}\\[2ex]
As may seen in Fig. 6, the quark model (even in its "relativized" versions !)
fails in describing the DHG. This is very surprising, indeed, because the model
gives a good overall description
of the excitation spectrum of the
nucleon (Isgur and Karl, 1978 and 1979). In the following we will demonstrate
the
reasons for this blatant
failure for the case of its nonrelativistic version. In its simplest version
the
model has a quark mass $m_q = m/3$, an oscillator parameter related to the
size of the object, $\alpha_0^{-2} = < r^2 >$,
and Dirac point particles leading to $\kappa= 2$. Including the usual
hyperfine interaction and a configuration mixing of
$0 \hbar \omega$ and $ 2 \hbar \omega$
states, the wave function of the nucleon is
\begin{equation}
| N \rangle = a_S | ^2 S_{1/2} \rangle_S + a_{S'} | ^2 {S'}_{1/2}
\rangle_S + a_M | ^2 {S'}_{1/2} \rangle_M + a_D | ^4 D_{1/2} \rangle_M,
\end{equation}
with admixture coefficients $a_S = 0.93, a_{S'} = - 0.29,
a_M = - 0.23$, and $a_D = - 0.04$ (Giannini, 1991).

The corresponding strength of the hyperfine interaction has been obtained by
fitting the positions of the first and second resonance region. The final
result
for the DHG integral is (Drechsel and Giannini, 1993; DeSanctis {\it et al.},
1994)
\begin{equation}
I( Q^2 = 0 ) = -1 +2 a_M^2 +\frac{5}{2} a_D^2 \pm \frac{1}{2} a_D^2
+ [ a^4 ] ,
\end{equation}
the upper and lower sign corresponding to proton and neutron, respectively. A
comparison with eq. (13) gives $I_{VV} = - 0.86$, in excellent agreement
with experiment, and $I_{SV}$ with the proper sign but
too small in magnitude. The
isoscalar magnetic moment cannot be explained by the small $D$-state admixture,
but probably requires an introduction of sea quark effects as in the case of
the
Ellis-Jaffe sum rule (Ellis and Jaffe, 1974). In
order to obtain the result of eq. (37) independently from
both the integral (sum over the excited states) and the ground state value of
the magnetic moment, the calculation has to be performed very "carefully",
however. In fact, the complete calculation without the hyperfine interaction
gives
\begin{equation}
I = -1 + \frac{5}{4} \frac{1 + \tau_0}{2} \sum_{ n \ge 1} \frac{1}{n !}
( n \zeta )^{ 2 n} e^{-(n \zeta)^2} ,
\end{equation}

where the sum is over all oscillator shells $(n\ge 1)$, and the expansion
parameter is
\begin{equation}
\zeta^2 = ( \hbar \omega_0 / \sqrt{3} \alpha_0 )^2 = 1 / ( 3 m_q^2 <r^2> )
\approx 0.57 ,
\end{equation}
with a value for $\alpha_0$ to describe the helicity structure of the spectrum
(Copley {\it et al.}, 1969).
Clearly the higher order terms are retardation terms of higher
order in $(\omega/m_q)^2$ than can be described in a nonrelativistic model. The
correct result can only be obtained for the leading order term, i.e. by
neglecting all terms of order $m_q^{-2}$ or, alternatively, by replacing
the $\omega^2$-dependence of the
retardation terms by a relativistic $Q^2$-dependence.

Even if we neglect higher order retardation, a further inconsistency appears if
the hyperfine interaction is switched on. The reason has been pointed out long
ago (Brodsky and Primack, 1969; Close and Copley, 1970; Krajcik
and Foldy, 1974; DeSanctis and Prosperi, 1987).
In order to fulfill the algebra of the Poincar\'e group (translations,
rotations and boosts)
for an interacting many-body
system, two-body currents are required,
\begin{equation}
\vec J = \sum_k \left( \frac{ \vec p_k}{m_q} + i \frac{\vec \sigma_k \times
\vec q }{2 m_q} \right) e^{i \vec q \cdot \vec r_k}
+ \vec J_{rel} ( \mbox{1-body} ) + \vec J_{rel} ( \mbox{2-body} ) .
\end{equation}

It is not sufficient to include the relativistic spin-orbit current and other
corrections of order $m_q^{-2}$ as in the usual "relativized"
versions of the CQM.
Instead, genuine two-body currents of order $m^{-1} m_q^{-1}$
appear at the same
level, in particular a modification of the electric dipole current due to $cm$
correlations of the relativistic system. Being functions of the properties of
both the struck particle and the total system (total charge, mass and momentum)
they are somewhat difficult to treat and, certainly, have been ignored within
the framework of single particle transitions.

With the current operator (40) and the given spectrum of the CQM, the sum rule
is
\begin{equation}
I \sim \sum_f \frac{1}{\omega^2} \left| \langle f | J_+ | i \rangle
\right|^2_{ \cal A - \cal P} ,
\end{equation}
where $\cal A - \cal P$
denotes the difference of the matrix elements for antiparallel spins
(initial state nucleon:$- 1/2$, photon:$+ 1$) and parallel spins $(+ 1/2,
+1)$. Eq. (41) can be expressed in terms of a vector product of the current
operators,
\begin{equation}
I \sim \sum_f \frac{1}{\omega^2} \left[ \langle i | \vec J^+ | f \rangle
\times \langle f | \vec J | i \rangle \right]_z,
\end{equation}
the excitation energy being a function of the states, $\omega = \omega_{fi}$.
Apparently the leading order convection current does not contribute,
because the
$\vec{p} \times \vec{p}$ contributions vanish identically. The DHG is saturated
by
the spin current $(\sim\omega^2\vec{\sigma} \times
\vec{\sigma})$, corresponding
retardation terms in the orbital angular momentum
$(\sim\omega^2\vec{l} \times \vec{l})$
and relativistic corrections of both one-body
and two-body structure of order $\omega^2$. Neglecting higher order terms
$O (\omega^4)$, the $\omega$-dependence in eq. (42)
cancels and the DHG integral may be
evaluated by closure.

Such an evaluation by closure can also be obtained for the individual multipole
contributions (Drechsel and Giannini, 1993; DeSanctis {\it et al.}, 1994).
The main results are
\begin{itemize}
\item the convection current cancels to leading order,
\item the remaining contribution of the $E_{0+}$-multipole is cancelled by the
complex of $E_{2-}/M_{2-}$ excitations,
\item up to relativistic corrections
and small contributions of the hyperfine force,
  only the (unretarded) spin-part of the $M_{1+}$ survives,
\item the contributions of $4 \hbar \omega_0$-
states to $I_{SV}$ seem to be large,
\item good agreement is reached for $I_{VV}$, while the predicted value of
$I_{SV}$ is much too small.
\end{itemize}

Along these lines we obtain a phenomenological prediction for $Q^2 > 0$ by
replacing
\begin{equation}
J_+ = \frac{\omega \sigma_+}{2 m_q} e^{i \vec q \cdot \vec r} \Rightarrow
\frac{\omega \sigma_+}{2 m_q} \left( 1 - \frac{Q^2 < r^2 >}{6} \right)
+ \frac{\sigma_+}{2 m_q} \frac{Q^2}{\omega + | \vec q |} .
\end{equation}
The first term on the $rhs$ shows the unretarded spin current multiplied by a
typical form factor, leading to a decrease in absolute value with increasing
$Q^2$. It is superimposed with the second term, carrying the same sign as for
the case of $Q^2= 0$. As a result the slope of the sum rule at $Q^2 = 0$
could be both positive or negative, depending on the form factor.\\[2ex]
\underline{Chiral Perturbation Theory (ChPT)}\\[2ex]
The discussion of the slope of the DHG integral has been reactivated by a
recent
calculation in ChPT (Bernard {\it et al.}, 1993). While a calculation of the
integral itself,
being of order $\kappa^2$, would require at least a two-loop calculation, its
derivative has been obtained both in the framework of relativistic ChPT and
within the heavy baryon approximation. The result is shown in Fig. 8. Obviously
the difference between the two predictions is large, and both differ from the
result of the "phenomenological" prediction (Fig. 6). The wide range of the
theoretical predictions is connected with the bad convergence of the loop
expansion in the case of the nucleon. Contrary to pionic problems, where all
energies, momenta and masses are small near threshold, the large mass of the
nucleon sets an additional (large !) scale. As a consequence, the explicit
$1/m$
expansion of the heavy baryon formalism converges much faster and, probably,
leads to a better prediction. As shown in the previous subsection, however,
relativistic corrections play an important role in the case of the DHG, and it
remains to be seen whether the leading term in the heavy mass formulation of
the
ChPT is really sufficient. Though $\Delta$-loops do not play a major
role at the one-loop level, it is questionable whether such resonance phenomena
can be appropriately described to that order.\\[2ex]

\underline{Current Algebra and large $Q^2$}\\[2ex]
At higher values of $Q^2$ an expansion of the current in $m_q^{-2}$ does not
make
sense. Instead one has to use the relativistic current operator. A "back of an
envelope" calculation  gives for the relevant component of the current, in the
Breit frame of the struck parton,

\begin{equation}
J_+ = \sum_{k}\frac{e_k\mid\vec{q}\mid}{2m_q}\sigma_+(k),
\end{equation}
where $Q^2 =\vec{q^2}$ in the Breit system. The DHG integral, eq. (41), becomes
\begin{equation}
I \sim \sum_f \sum_k \frac{1}{\nu^2} \left| \left\langle f \left|
\frac{e_k\mid\vec{q}\mid}{2m_q}\sigma_+(k) \right| i \right\rangle
\right|^2_{\cal A -\cal P } .
\end{equation}
Rewriting this expression in terms of the Bjorken scaling variable $x$, we
obtain
\begin{equation}
I \sim \frac{1}{Q^2} \sum_k \sum_f \left| \left\langle f \left|
e_k \frac{m}{m_q} x \sigma_+(k) \right| i \right\rangle
\right|^2_{\cal A -\cal P } .
\end{equation}
In the naive parton model we find $x=m_q/m \approx 1/3$. Hence the
sum over the final states may be performed, and
\begin{equation}
I ( Q^2 ) = \frac{2 m^2}{Q^2} \Gamma_1, \qquad \Gamma_1 = \frac{1}{6}
\cdot \frac{5}{3} \cdot \frac{1 + \tau_0}{2} .
\end{equation}
The factor $g_A/g_V = 5/3$ is the prediction of the simple quark model for the
axial coupling constant, and $\Gamma$ is related to the quark spin distribution
by
\begin{equation}
I (Q^2) \Rightarrow \frac{2 m^2}{Q^2} \int_0^1 dx g_1(x) \equiv
\frac{ 2 m^2}{Q^2} \Gamma_1 .
\end{equation}
We note in passing that
\begin{equation}
\Gamma_1^p - \Gamma_1^n = \frac{1}{6} \frac{g_A}{g_V} ( 1 - \alpha_s ( Q^2 )
\pm \ldots ) \approx 0.19
\end{equation}
is given by current algebra (Bjorken, 1966), while the individual values of
$\Gamma^p$ and $\Gamma^n$ become model dependent
(Ellis and Jaffe, 1974). For a more detailed discussion see the contribution by
B. Frois
(1994). As has been pointed out in sect. 3, the Burkhardt-Cottingham sum
rule implies $\Gamma_2^p = \Gamma_2^n = 0$ for $Q^2\rightarrow\infty$. \\

Let us finally comment on the role of current algebra for the DHG sum rule. The
asymptotic limit of the sum rule at large $Q^2$ was first discussed by Bjorken
(1966) on the basis of the equal-time current commutator
\begin{equation}
[j_\mu ( 0, \vec r ) , j_\nu ( 0, 0) ] = -2 i \epsilon_{ \mu \nu
\lambda \rho} q^\lambda j_5^\rho \delta(\vec r) + \mbox{gradients}.
\end{equation}
Comparing with our eq. (42), we immediately find that the vector product in
that
equation is nothing else than the commutator of the space-like parts of the
current, hence $(\vec{J} \times \vec{J})_z \sim (\vec{J}_5)_z = g_A \bar{\psi}
\gamma_z \gamma_5 \psi $.
As a result the DHG integral in the scaling region is given by the axial
current, in particular $I^p-I^n \sim g_A^{(3)} \approx  5/4$.
In view of the experimental evidence in 1966, Bjorken was not too much
impressed
with the possible consequences of his work. He
wrote: "Something has to be salvaged
from this worthless equation by constructing an inequality...", and derived an
upper limit for the spin-averaged total cross section.\\[2ex]
\underline{QCD Sum Rules}\\[2ex]
As an example of higher-twist calculations extrapolating Bjorken's result to
smaller $Q^2$, we refer to a recent QCD based prediction of Balitsky {\it et
al.}(1990). They find only small corrections
to the asymptotic behaviour,
\begin{equation}
\Gamma^p - \Gamma^n \approx \frac{1}{6} \left( (1-\alpha_s(Q^2)) g_A -
\frac{0.3 GeV^2}{Q^2} + \left[\frac{1}{Q^4}\right] \right) .
\end{equation}
Even at the $Q^2 = 1 GeV^2$, the smallest reasonable value for such an
expansion, the correction is only 25\%. Hence the DHG integral at
the real photon point should be saturated by contributions
dying out faster than
$1/Q^4$ in the asymptotic limit.\\[2ex]

\hspace{2.5cm}THE SPIN STRUCTURE OF RESONANCES\\[2ex]
The integrand of the DHG sum rule is determined by the helicity structure of
the
integrated cross section. In the resonance region these contributions may be
decomposed either in electric and magnetic multipoles or in "helicity
amplitudes",
\begin{eqnarray}
A_{1/2} & = & \sqrt{\frac{4 \pi \alpha}{2 k_{\gamma}^{cm}}}
\langle N^* ( J', M'= + \frac{1}{2} ) \mid
J_{+} \mid N^* ( J = \frac{1}{2}, M= - \frac{1}{2} ) \rangle\\
A_{3/2} & = & \sqrt{\frac{4 \pi \alpha}{2 k_{\gamma}^{cm}}}
\langle N^* ( J', M'= + \frac{3}{2} ) \mid
J_{+} \mid N^* ( J = \frac{1}{2}, M= + \frac{1}{2} ) \rangle\\
S_{1/2} & = & \sqrt{\frac{4 \pi \alpha}{2 k_{\gamma}^{cm}}}
\langle N^* ( J', M'= + \frac{1}{2} ) \mid
J_{+} \mid N^* ( J = \frac{1}{2}, M= + \frac{1}{2} ) \rangle,
\end{eqnarray}
the latter describing the longitudinal current.
In the case of the $A_{\frac{3}{2}}$ amplitude, the photon can be absorbed by a
single quark without a helicity flip, while the $A_{\frac{3}{2}}$ amplitude
requires quark spin flips (see Fig. 9). Since the quark masses can be neglected
in the limit of large momentum transfer, $A_{\frac{1}{2}}\sim Q^{-3}$ becomes
the
dominant amplitude in that limit, and $A_{\frac{3}{2}}\sim Q^{-5}$ should be
strongly suppressed (LePage and Brodsky, 1980)\\[2ex]

\underline{The First Resonance Region}\\[2ex]
This region between threshold and about $400 MeV$ excitation energy is
dominated
by the $P_{33}$ (1232) or $\Delta$(3,3) resonance, clearly visible in Fig. 3 on
top of a broad background of mostly S-wave pions. Within the harmonic
oscillator
quark model, the $\Delta$ and the nucleon are partners with configuration
$\{56,0^+\}_0$,
i.e. members of the symmetrical 56-plet of $SU(6)$, orbital momentum $L
= 0$, positive parity and no radial nodes. In this approximation the $\Delta$
may only be excited by the magnetic dipole ($M1$ or $M_{1+}$, respectively).
As has been stated previously, the introduction of a hyperfine interaction
leads
to an admixture of mixed symmetry states of the 70-plet in connection with
orbital or radial excitation. Of particular significance is the admixture of a
$D$-state component leading to the existence of a small electric quadrupole
transition ($E2$ or $E_{1+}$, respectively). The helicity amplitudes for this
resonance are superpositions of the corresponding multipoles,
\begin{equation}
A_{\frac{1}{2}} = - \frac{1}{2} (M1 + 3 \cdot E2),\quad A_{\frac{3}{2}} = -
\frac{\sqrt{3}}{2} (M1 - E2).
\end{equation}
Without hyperfine interactions, these amplitudes are simply proportional,
$A_{\frac{3}{2}} = \sqrt{3} A_{\frac{1}{2}}$. On the other hand, perturbative
$QCD$ predicts that the spin-flip amplitude $A_{\frac{3}{2}}$ should vanish.
Therefore the ratio $EMR \equiv E2/M1$ should approach unity in the limit $Q^2
\rightarrow\infty$.
In the low-energy regime, however, the D-state probability of both nucleon and
$\Delta$ is of the order of 1 $\%$,
leading only to a small quadrupole moment of
the $\Delta,\,Q_{\Delta} \approx -0.089fm^2.$

Careful studies have shown that the
polarized photon asymmetry $\sum$ is the most sensitive observable for
experiments with real photons (Blanpied {\it et al.}, 1992). Many
more choices seem to exist for
electroexcitation with polarization degrees of freedom, apparently some of the
longitudinal and transverse interference terms are very sensitive to both the
$E_{1+}$ and $L_{1+}$ amplitudes. The coincidence $\vec{e} + p \rightarrow e' +
\vec{p'} + \pi^0$, with polarization transfer to the proton, is a particularly
well suited experiment (Lourie, 1990; Hanstein, 1993).
The present value is $ EMR \approx - 1.5 \% $
at the real photon point
with some indications that it becomes slightly positive at
$ Q^2 \approx 3GeV^2 $ (see Fig. 10).
The corresponding ratio $SMR \equiv S_{1+}/M_{1+}$ is also negative with large
error bars and partially contradicting experimental evidence. The recent Bonn
data (Kalleicher, 1993) indicate a relatively strong
fluctuation as function of $Q^2$. The
measured value at $Q^2\approx 0.1 GeV^2$, $SMR \approx - 13 \% $,
corresponds
to $EMR \approx - 6\% $.
\\[2ex]

\underline{The Roper Resonance $P_{11}$(1440)}\\[2ex]
In the $CQM$ the Roper is a radial excitation of the nucleon occuring at an
energy of $2\hbar\omega_0$. In units of $10^3/GeV^\frac{1}{2}$, the measured
helicity amplitude for the proton is $A^p_{\frac{1}{2}}= -70\pm 5$, the value
for the neutron is $23\le A^n_{\frac{1}{2}} \le 56$. The $CQM$ predicts a ratio
$A^n_{\frac{1}{2}}/A^p_{\frac{1}{2}} = - \frac{2}{3}$, in reasonable agreement
with the data within the large error bars. However, the values for the $CQM$
amplitudes themselves are too small by a factor of 3. The chiral bag model
($CBM$) predicts a ratio of $-1$ for the pionic contributions. With
decreasing bag
radius $r_0$, these effects of the pion cloud increase strongly, e.g.
$A^p_{\frac{1}{2}} = -36, -80 $ and $-147$ for $r_0=1fm, 0.8fm$ and $0.6fm$,
respectively (Drechsel, 1994).

As has been pointed out by Li {\it et al.} (1992),
explicit gluon degrees of freedom might play a role even at low excitation
energies. The wave function of such a "hybrid" contains two components,
\begin{equation}
\mid N^* \rangle = \alpha\mid q^3 \rangle +\sqrt{1-\alpha^2\mid} q^3
\times g \rangle.
\end{equation}
The gluon appearing in the second term requires a quark configuration
$q^3$ with colour in order to insure an overall colour neutral wave function.
As
a consequence the quarks can now be in the configuration $[70, 0^+]_0$, i.e.
with mixed symmetry in $SU$ (6) classification and neither orbital nor radial
nodes. In this case the wave function in $\vec{r}$-space may be identical to
that one of the nucleon, leading to
\begin{equation}
A_{\frac{1}{2}} (q^3 g)/A_{\frac{1}{2}} (q^3) \sim \frac{1}{Q^2}.
\end{equation}
The corresponding Coulomb amplitude vanishes except for relativistic
corrections, $S_{\frac{1}{2}} (q^3 g) \approx 0$, because
the longitudinal photon
cannot excite the transverse colourmagnetic field of the gluon. The Roper is
certainly a good candidate for such a "hybrid", because it occurs at an
extremely low energy for a $2\hbar\omega_0$ state of the $CQM$. Up to now the
Roper has not been seen very clearly in electromagnetic reactions. The size of
its Coulomb excitation $S_{\frac{1}{2}}$ will be quite essential for its
classification. While a small or vanishing value will be an indication of a
hybrid, very large contributions should be typical of explicit pion degrees of
freedom as predicted by the $CBM$. The present status of the data on the Roper
is compared to various predictions in Fig. 11.

It is also interesting to note that a broad bump has been seen near the Roper
resonance in a missing energy spectrum for $\alpha - p$ scattering, which
could be an indication for a strong monopole transition (Morsch {\it et al.},
1992).\\[2ex]

\underline{The Second and Third Resonance Region}\\[2ex]
The second resonance region contains the two dipole excitations
$S_{11}$(1535) and
$D_{13}$(1520) with spins $\frac{1}{2}^-$ and $\frac{3}{2}^-$, respectively. In
the $CQM$ its configurations are $\{70_M, 1^-_M\}_1$. The most prominent state
in
the third resonance region is the $F_{15}$(1680) with configuration
$\{56_S,2^+\}_2$. These states have quite different properties as function of
momentum
transfer. In the $CQM$ we have
\begin{eqnarray}
A_{\frac{1}{2}} (S_{11}) & \sim & \left( \frac{\vec{q\,}^2}{\alpha^2} +
2 \right) F (\vec{q\,}^2), \nonumber \\
A_{\frac{1}{2}} (D_{13}) & \sim & \left( \frac{\vec{q\,}^2}{\alpha^2} -
1 \right) F (\vec{q\,}^2), \nonumber \\
A_{\frac{3}{2}} (S_{11}) & \sim & F (\vec{q\,}^2), \\
A_{\frac{1}{2}} (F_{15}) & \sim & \left( \frac{\vec{q\,}^2}{\alpha^2} -
2 \right) \mid \vec q \mid F (\vec{q\,}^2), \nonumber \\
A_{\frac{3}{2}} (F_{15}) & \sim & \mid \vec q \mid F (\vec{q\,}^2). \nonumber
\end{eqnarray}
The two contributions to the $A_{\frac{1}{2}}$ amplitudes are due to the spin
and orbital currents of the quark motion.
At the real photon point, $\vec{q\,}^2 =
\omega^2$, the experiments indicate a cancellation of the two currents in the
case of the proton, $A^p_{\frac{1}{2}} (D_{13}) \approx 0\approx
A^p_{\frac{1}{2}}(F_{15})$. Using $\alpha\sim 0.17 GeV^2$, this cancellation is
nearly complete for both resonances, which may be considered as one of the
early
successes of the $CQM$ (Copley {\it et al.}, 1969). Replacing
$\vec{q\,}^2\rightarrow Q^2$ (i.e.
performing the nonrelativistic calculations in the Breit frame), we find that
the amplitudes $A_{\frac{1}{2}}$ become increasingly important for large
$Q^2$.
This is in agreement with $PQCD$, requiring the dominance of the helicity
conserving amplitudes $A_{\frac{1}{2}}$ in the asymptotic region. Within
$PQCD$
the prediction is $A_{\frac{1}{2}} \sim Q^{-3}$ and $A_{\frac{3}{2}} \sim
Q^{-5}$
for $Q^2 \rightarrow\infty$. The rapid change at
$Q^2\approx 0.5GeV^2$ is reflected most clearly by the helicity asymmetry shown
in Fig. 12. The value $(A_{\frac{1}{2}} - A_{\frac{3}{2}})/(A_{\frac{1}{2}} +
A_{\frac{3}{2}})$ ranges between the lowest possible ratio -1 at the real
photon
point and the highest possible ratio +1 for $Q^2 \rightarrow\infty$, for the
strongest states of both the second $(D_{13})$ and third $(F_{15})$ resonance
region.\\[2ex]

\underline{Eta Production}\\[2ex]
In comparison with its partner $D_{13}$(1520), the dipole excitation
$S_{11}$(1535) is only weakly seen in pion photoproduction. However, it couples
very strongly to the $\eta$ meson, about 50\% of its decay width is due to
$\eta$
emission. In comparison, the $D_{13}$ has only a $10^{-3}$ branch for $\eta$
decay, and also an excited $S_{11}$ occuring in the third resonance region
couples only weakly to the $\eta$. The only other resonance with a sizeable
$\eta$ branch is the $P_{11}$(1710) with 25\% $\eta$ decay. Though the overall
contribution of the $\eta$ to sum rules will be small, the study of this decay
channel is interesting because of its connection with strangeness degrees of
freedom. The data seem to indicate a rather slow decrease of the transition
form
factor to the $S_{11}$ resonance as function of $Q^2$.\\[2ex]

\hspace{2.5cm}{PERSPECTIVES AND CONCLUSIONS}\\[2ex]
Investigations with electromagnetic interactions have contributed substantially
to a better understanding of the structure of hadrons. However, previous
experiments have been limited by small currents and low duty-factors. As a
consequence the statistics for small amplitudes has been bad and the signal to
noise ratio has been small. With the advent of the new electron accelerators
new
classes of coincidence experiments have become possible, and polarization
degrees of freedom will play an important role. With a beam polarization of
40\%
and more, polarized electrons promise to provide a new capability to measure
some of the most wanted observables, in particular in combination with target
and recoil polarization.
In the nucleon resonance region such systematic investigations with complete
kinematics and separation of the independent structure functions include:
\begin{itemize}
\item a model independent determination of the quadrupole amplitudes $E_{1+}$
and $L_{1+}$ in the region of the $\Delta$ resonance ("bag deformation"),
\item the measurement of the monopole strength $L_{1-}$ near the Roper
resonance
("breathing mode vs. hybrid"),
\item the analysis of the helicity asymmetry of the nucleon resonances with its
strong dependence on momentum transfer,
\item the "tagging" of the weak $S_{11}$ dipole resonance by the $\eta$ channel
and, by precision experiments, the coupling of the $\eta$ to other resonances
as
function of momentum transfer $Q^2$.
\end{itemize}
The helicity structure of the photo- and electroproduction cross sections is
related to the spin structure of the nucleon in deep inelastic lepton
scattering. Both the generalized Drell-Hearn-Gerasimov sum rule and the
Burkhardt-Cottingham sum rule define energy-weighted integrals over the
excitation spectrum from the photonuclear point $(Q^2=0)$ to asymptotic values
of momentum transfer, where the experiment probes the spin
distribution function. Since these sum rules have been derived on the basis of
quite
general principles (relativity, causality, unitarity, gauge invariance), they
provide a unique testing ground for our understanding of the nucleon. In
particular, the sum rules connect ground state properties (magnetic moments and
form factors) with the helicity structure of the excitation spectrum.

Up to now neither of these sum rules has been tested by a direct experiment.
There is
still the possibility that the sum rules will not converge. Such a failure
would
indicate that even the ground state properties of the nucleon are determined by
phenomena happening at asymptotically large energies, a situation which would
send all model-builders back to the drawing board.
A series of experiments is underway to clarify the situation. In a
collaboration
of Bonn and Mainz groups (Arends {\it et al.}, 1993), the spin structure in the
resonance region will be
studied with real photons to find out whether the DHG sum rule converges and,
ultimately, whether the proton-neutron difference is an indication of a
possible
breakdown of our theoretical concepts.

In the region of the order of $Q^2\sim 1 - 2 GeV^2$, various CEBAF experiments
(Burkert {\it et al.}, 1991; Kuhn {\it et al.}, 1993) will explore the sum
rules in the transition
region from coherent resonance excitation to deep inelastic scattering. Of
particular interest will be the rapid crossover of the DHG integral from large
negative to positive values and the question whether the predictions of the
Burkhardt-Cottingham sum rule can be established.
All of these experiments will require a high degree of precision and a careful
analysis of the systematic errors. However, they will help to increase our
knowledge of hadronic structure in a truly qualitative way and provide a good
chance to discover new and exiting phenomena.

\vspace{5ex}
\hspace{2.5cm}REFERENCES\\[2ex]
Anselmino, M., B.L. Ioffe and E. Leader (1989), Sov. J. Nucl. Phys. {\bf 49},
136.\\
Arends, J. {\it et al.} (1993), Proposal to measure the
Gerasimov-Drell-Hearn sum rule, Bonn and Mainz.\\
Balitsky, I.I., V.M. Braun and A.V. Kolesnichenko (1990), Phys. Lett.
{\bf B242}, 245.\\
Bernard, V., N. Kaiser and U.-G. Meissner (1993), Phys. Rev.
{\bf D48}, 3062.\\
Bjorken, J.D. (1966), Phys. Rev. {\bf 148}, 1467.\\
Blanpied, G.S. {\it et al.} (1992), Phys. Rev. Lett. {\bf 69}, 1880.\\
Brodsky, S.J. and J. Primack (1969), Ann. Phys. (N.Y.) {\bf 52}, 315.\\
Burkert, V. (1990), Proc. Workshop on the Hadron Mass Spectrum,
Rheinfels.\\
Burkert, V.D. {\it et al.} (1991), Measurement of polarized structure
functions in inelastic electron scattering using CLAS, CEBAF-PR-{\bf
91-023}.\\
Burkert, V.D. and B.L. Ioffe (1992), Phys. Lett. {\bf B296}, 223.\\
Burkhardt, H. and W.N. Cottingham (1970), Ann. Phys. (N.Y.) {\bf 56},
543.\\
Chang, L.N., Y. Liang and R. Workman (1992), University of Rochester
Report No. UR-1258 (unpublished).\\
Chew, G.F. {\it et al.} (1957), Phys. Rev. {\bf 106}, 1345. \\
Close, F.E. and L.A. Copley (1970), Nucl. Phys. {\bf B19}, 477.\\
Copley, L.A., G. Karl and E. Obryk (1969), Nucl. Phys.
{\bf B13}, 303 and Phys. Lett. {\bf 29B}, 117.\\
DeSanctis, M. and D. Prosperi (1987), Il Nuovo Cim. {\bf 98A}, 621. \\
DeSanctis, M., D. Drechsel and M.M. Giannini (1994), Few-Body-Systems {\bf 16},
143\\
Drechsel, D. and L. Tiator (1992), J. Phys. G: Nucl. Part.
Phys. {\bf 18}, 449.\\
Drechsel, D. and M. M. Giannini (1993), Few-Body Systems {\bf 15}, 99.\\
Drechsel, D. (1994), Few-Body Systems, Suppl.{\bf7}, 325.\\
Drell, S.D. and A. C. Hearn (1966), Phys. Rev. Lett. {\bf 16}, 908. \\
Ellis, J. and R. Jaffe (1974), Phys. Rev. {\bf D9}, 1444 and {\bf D10},
1669.\\
Feynman, R.P. (1972), Photon-Hadron Interactions, Benjamin (Reading, MA).\\
Frois, B. (1994), Proc. Int. School of Nuclear Physics, Erice, to be published
by Pergamon Press (Oxford)\\
Gell-Mann, M. and M.L. Goldberger (1954), Phys. Rev. {\bf 96}, 1433.\\
Gerasimov, S.B. (1966), Sov. J. Nucl. Phys. {\bf 2}, 430.\\
Giannini, M. M. (1991), Rep. Progr. Phys. {\bf 54}, 453.\\
Hanstein, O. (1993), diploma thesis, Mainz.\\
Heimann, R.L. (1973), Nucl. Phys. {\bf B64}, 429.\\
Isgur, N. and G. Karl (1978 and 1979), Phys. Rev. {\bf D18}, 4187 and
{\bf D19}, 2653.\\
Kalleicher, F. (1993), Ph.D. thesis, Mainz.\\
Karliner, I. (1973), Phys. Rev. {\bf D7}, 2717.\\
Klein, O. and Y. Nishina (1929), Z. Physik {\bf 52}, 853.\\
Krajcik, R.A. and L. L. Foldy (1974), Phys. Rev. {\bf D10}, 1777.\\
Kuhn, S.E. {\it et al.} (1993), The polarized structure function $G_{1n}$
and the $Q^2$ dependence of the Gerasimov-Drell-Hearn sum rule for the neutron,
CEBAF-PR-{\bf 93-009}. \\
LePage, G.P. and S.J. Brodsky (1980), Phys. Rev. {\bf D22},
2157.\\
Li, Z.P., V. Burkert and Zh. Li (1992), Phys. Rev. {\bf
D46}, 70.\\
Lourie, R.W. (1990), Nucl. Phys. {\bf A509}, 653.\\
Low, F. (1954), Phys. Rev. {\bf 96}, 1428.\\
Morsch, H.P. {\it et al.} (1992), International Nuclear Physics
Conference, Contribution {\bf 2.1.17}, Wiesbaden.\\
Powell, J.L. (1949), Phys. Rev. {\bf 75}, 32.\\
Schwinger, J. (1975), Proc. Natl. Acad. Sci. USA {\bf 72}, 1.\\
Soffer, J. and O. Teryaev (1993), Phys. Rev. Lett. {\bf 22}, 3373.\\
Tsai, W.-Y., L. Deraad Jr. and K.A. Milton (1975), Phys. Rev. {\bf D 11},
3537.\\
Watson, K.M. (1954), Phys. Rev. {\bf 95}, 228.\\
Workman, R.L. and R.A. Arndt (1992), Phys. Rev. {\bf D45}, 1789.\\

\newpage

\vspace{5ex}
\hspace{2.5cm}FIGURES\\[2ex]
{Fig. 1:
Real or virtual Compton scattering off the nucleon. The
four-momenta of photon and nucleon in the initial state are denoted by
$q=(\omega,\vec{q})$ and $p = (E,\vec{p})$, respectively, with an additional
"prime" for
the final states. The photon polarizations are $\hat{\epsilon}$ and
$\hat{\epsilon}'$, and the
nucleon has charge $e$, mass $m$, and anomalous magnetic moment $\kappa$.
Note:  $\omega_{lab} = \nu.$}

{Fig. 2:
Measurement of the DHG sum rule. Left: the spins of
photon and nucleon are parallel, the projection of total angular momentum is
$J_z=\frac{3}{2}$. Right: antiparallel spins, $J_z=\frac{1}{2}$.}

{Fig. 3:
The total photoabsorption cross section $\sigma_T$ for the
proton in the resonance region as function of the photon energy $E_\gamma=\nu$.
Also shown are the main decay channels.}

{Fig. 4:
The difference of the photoabsorption cross sections for the two
helicities $\sigma_{3/2}- \sigma_{1/2}$ as function of $\nu$, in
the resonance region (Karliner, 1973). From left to right: isovector (VV),
isovector-isoscalar interference (SV), and isoscalar (SS) contributions.
Note the difference in scale !}

{Fig. 5:
Kinematics for double-polarization experiments. Left: The
incoming electron with helicity $h$ is scattered off a nucleon target with
polarization $\vec{P}$. The latter is analyzed in a frame with axes $x$ and
$z$ in
the electron scattering plane, and $y$ perpendicular to the plane. Right: The
recoil polarization of the nucleon is analyzed in the reaction plane of the
final-state hadrons, e.g. proton and pion. Its axes are $\vec{l}$ (along the
direction of the nucleon), $\vec{t}$ (sideways, or transverse in the reaction
plane), and $\vec{n}$ (perpendicular to the reaction plane).}

{Fig. 6:
The DHG integral for the proton, $I_p$, as function of $Q^2$
compared to different models. The full and dash-dotted line are phenomenlogical
models with different assumptions on the Roper resonance, the dashed and
double-dotted line
is the vector dominance model, the two dashed lines starting near the origin
are
the predictions of (relativized) quark models. The dashed line at positive
values indicates the data of EMC/SLAC experiments, the error bars have the
predicted accuracy of the planned CEBAF experiment (Burkert {\it et al.},

{Fig. 7:
The DHG integral for the neutron, $I_n$, as function of $Q^2$
compared to different models. The full curve is based on an analysis of pion
photoproduction in the resonance region, the dotted and dashed curves are for
different assumptions on the Roper resonance (nonrelativistic quark model and
hybrid state containing explicit gluon degrees of freedom), the curve labeled
ChPT has the slope predicted by Bernard {\it et al.} (1993).
The error bars have
the predicted accuracy of the planned CEBAF experiment (Kuhn {\it et al.},
1993).}

{Fig. 8:
Momentum dependence $(-k^2=Q^2)$ of the extended DHG sum rule,
$\tilde{I}_p(Q^2) = I_p (Q^2)-I_p (0)$. The solid line gives the one-loop
result in the heavy baryon limit of ChPT, the dashed line includes additional
one-loop graphs with $\Delta (1232)$ resonances, and the dot-dashed curve is
the
result of the relativistic one-loop version of ChPT (Bernard {\it et al.},
1993).}

\newpage

{Fig. 9:
The "helicity amplitudes" for
electroproduction of nucleon resonances for
collinear reactions. $A_{\frac{1}{2}}$: Incoming nucleon with spin projection
$m= -\frac{1}{2}$ (positive helicity) absorbs a photon with spin $\lambda = +
1$,
leading to $m' = \frac{1}{2}$ (same helicity as in initial state).
$A_{\frac{3}{2}}:$ For initial spin $m=+\frac{1}{2}$ (negative helicity) and a
photon with $\lambda = +1$, the final spin is $m' = \frac{3}{2}$ (positive
helicity, helicity change). $S_{\frac{1}{2}}:$ For longitudinal photons
$(\lambda = 0)$ the conservation of spin requires a helicity change.}

{Fig. 10:
The ratio of the electric to magnetic multipole strength for
the $\Delta$(1232) resonance as function of momentum transfer. Left:$\mid
E_{1+}/M_{1+}\mid$, data from (Burkert, 1990). Right: $\mid S_{1+}/M_{1+}\mid$.
Note that the new data point from Bonn $(\bullet)$ at $Q^2\approx. 12 GeV^2$
translates into a value of about - 6$\%$ for the ratio
$\mid L_{1+}/M_{1+}\mid$
(Kalleicher, 1993).}

{Fig. 11:
The amplitudes $A_{\frac{1}{2}}$ and $S_{\frac{1}{2}}$ as
function of $Q^2$ for the Roper resonance. Left: Long-dashed line for $q^3$
model, solid line $q^3 g$ state, other lines various data analyses. Right:
solid
line $q^3$ model, vanishing amplitude for $g^3 g$ state, other lines various
data analyses (Li {\it et al.}, 1992)}.

{Fig. 12:
The helicity asymmetry $(A_{\frac{3}{2}} -
A_{\frac{1}{2}})/(A_{\frac{3}{2}}+ A_{\frac{1}{2}})$
for electroexcitation of the $D_{13}$(1520) and the $F_{15}$(1680)
as function of $Q^2$. The data have been
compared to various quark model calculations (Burkert, 1990)}.

\end{document}